\documentclass[twocolumn]{article} 
\usepackage[utf8]{inputenc}
\usepackage{amsmath}
\usepackage{graphicx}
\usepackage[a4paper, total={6in, 8in}]{geometry}
\usepackage{listings}
\usepackage{float}
\usepackage{placeins}
\usepackage[nottoc,numbib]{tocbibind}
\usepackage{xcolor}
\usepackage{hyperref}
\hypersetup{colorlinks,linkcolor={blue},citecolor={blue},urlcolor={blue}}
\usepackage[]{lineno}
\usepackage{bm}
\usepackage{url}
\usepackage[affil-it]{authblk}

\usepackage[style=phys]{biblatex}
\title{An Electromagnetic Particle-Particle Model
on Solving Relativistic Binary Collision}

\author[1,2]{Yanan Zhang}
\author[2]{Xiaochun Ma}
\author[1]{Hui Liu}
\author[1,*]{Yinjian Zhao}

\affil[1]{School of Energy Science and Engineering,
Harbin Institute of Technology,
Harbin 150001, People’s Republic of China}
\affil[2]{Harbin Boiler Company Limited,
Harbin 150040, People’s Republic of China}
\affil[*]{Corresponding author: Yinjian Zhao, zhaoyinjian@hit.edu.cn}

\date{\today}

\begin{document}


\twocolumn[
  \begin{@twocolumnfalse}
    \maketitle
    \begin{abstract}

With the significant advancements in parallel computing techniques,
the particle-particle (PP) model has been effectively
utilized in various plasma-related applications.
However, PP has been limited for solving only electrostatic problems
under Coulomb's law,
by analogy to the particle-in-cell (PIC) model
solving Poisson's equation.
While electromagnetic PIC is common
with coupled solutions of Maxwell's equations,
we propose an electromagnetic (EM) PP model
taking advantage of Li\'enard-Wiechert potentials
for point charge in this paper.
In addition, this EM-PP model can contribute
to simulate relativistic binary collisions with
high accuracy, thus its results are used as a baseline
to compare with the classical Frankel's relativistic
scattering angle,
and the accuracy and applicable scope of
Frankel's formula are discussed.

    \end{abstract}
  \end{@twocolumnfalse}
]

\section{Introduction}

Although not very common,
there have been a number of existing applications using
the particle-particle (PP) model
in plasma simulations in recent years.
For example, PP can be used to
solve the dynamics among charged droplets
in electrospray thrusters in the field
of plasma propulsion \cite{zhao,BREDDAN2023106079};
PP has been applied to simulate
ion beam neutralization in a large vacuum region
with separated ion and electron sources
\cite{doi:10.2514/1.B36770};
and PP has advantages to describe
plasma systems with dominant Coulomb collisions
\cite{10.1063/1.5025581,ZHAO20172944,10.1063/1.5025431}.

Comparing to the commonly used particle-in-cell (PIC)
method,
although PP is known for its
large computation,
so the number of simulated particles
is usually much smaller than that of PIC,
PP has several merits over PIC:
(1) PP solves both the long-range and short-range
inter-particle forces with
high accuracy, while PIC ignores the
forces within the range of a cell.
(2) Since PP does not use mesh,
it does not have mesh related restrictions and limitations,
such as the mesh resolution issue and
the domain size limitation.
(3) PP can be easily implemented in 3D
without greatly increasing computational cost
that PIC does.
(4) PP can be used for solving non-neutral
problems, while PIC was intrinsically designed for
simulating
quasi-neutral plasmas with Debye screening effects.

Among relevant works mentioned above,
PP has only been implemented
in electrostatics, i.e., the considered force
field is the Coulomb's law,
which is equivalent to the Poisson's equation
in PIC.
For electromagnetic problems,
PIC can solve the Maxwell's equations,
and the equivalent formulas for
point charges
are the Li\'enard-Wiechert potentials
$V$ and $\bm{A}$ \cite{Griffiths}:
\begin{equation}
    V(\bm{r},t) = \dfrac{1}{4 \pi \varepsilon_0}
    \dfrac{q c}{(\eta c - \bm{\eta} \cdot \bm{v})},
\end{equation}
\begin{equation}
    \bm{A}(\bm{r},t) = \dfrac{\bm{v}}{c^2} V(\bm{r},t),
\end{equation}
where $\bm{v}$ is the velocity of a point charge
$q$
evaluated at the retarded time
(labeled as $t_r$ in this paper and
will be introduced more in Sec.\ref{sec:GE}),
$\bm{\eta}$ is the vector from
the retarded position to the field point $\bm{r}$,
$\varepsilon_0$ is the vacuum permittivity, and
$c$ is the speed of light.
In this work,
it is shown that the electrostatic PP model
can be extended to become electromagnetic (EM),
by applying the electric and magnetic fields
derived from the Li\'enard-Wiechert potentials,
and implicitly solving the retarded time.
This EM-PP model is thus capable of simulating
EM inter-particle forces and particle dynamics
accurately without a mesh.

As a first application of the EM-PP model,
the scenario of relativistic binary collision
is chosen in this paper, which is very suitable for
EM-PP,
because there are only two
charges in vacuum,
and EM-PIC is not able to
solve the short-range EM forces.
Under the non-relativistic condition,
the binary collision has been solved theoretically
with an analytical formula of the scattering angle
available.
Under the relativistic condition,
there seems to be only one classical reference
\cite{PhysRevA.20.2120}
provided by N. E. Frankel et al. back to 1979.
The relativistic version of the binary collision
scattering angle is derived based on
some assumptions such as simplified force
fields and the impulse approximation.
To date, Frankel's formula has been
applied in many plasma applications
\cite{10.1063/1.4742167,
10.1063/5.0190352,10.1063/5.0102919}
and famous simulation programs
such as Smilei \cite{DEROUILLAT2018351}
and WarpX \cite{warpx}, but it seems to lack
a validation or a discussion on the
range that Frankel's formula possesses high accuracy.

Therefore, in addition to introduce the
EM-PP model, the relativistic binary collision
is also studied in this paper,
and the accuracy and applicable scope of
Frankel's scattering angle formula are discussed too.
The paper is organized as follows,
the EM-PP model is described in
Sec.\ref{sec:Method};
a test under constant velocity
for validating the EM-PP model is given
in Sec.\ref{sec:3.1};
simulations on the relativistic binary collision
are presented in Sec.\ref{sec:3.2};
and conclusions are drawn in Sec.\ref{sec:conclusion}
at last.

\section{Method}\label{sec:Method}

\subsection{Governing Equations}\label{sec:GE}

\begin{figure}[!ht]
\centering
\includegraphics[width=0.3\textwidth]{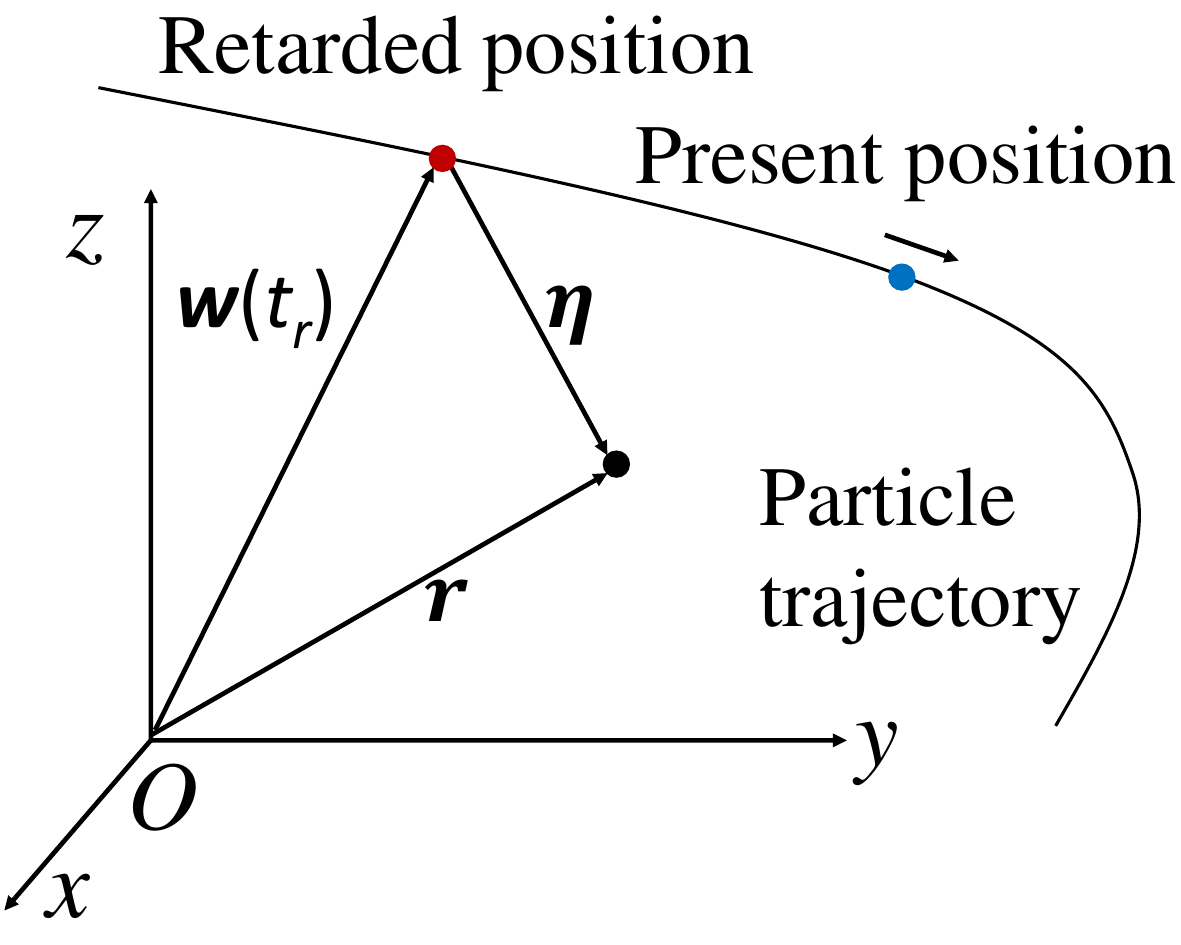}
\caption{
Illustration of the retarded time
and the vector $\bm{\eta}$.
}
\label{fig:EMP}
\end{figure}

Consider a point charge $q$ moving in vacuum,
its trajectory is a function of time $t$,
denoted by $\bm{w}(t)$.
While moving, the point charge generates
electric $\bm{E}$ and magnetic $\bm{B}$ fields
in space,
and the newly generated fields need to
act on another charge at position $\bm{r}$ with
a delay, because the fields are established
or propagate at the speed of light $c$.
Therefore, we define the retarded time $t_r$,
and it was at time $t_r$ when
the charge was at position $\bm{w}(t_r)$,
it generated the fields that
arrives at the position $\bm{r}$ in space
at the present time $t$,
as illustrated in Fig.\ref{fig:EMP}.
The retarded time thus satisfies the relation:
\begin{equation}
    |\bm{r} - \bm{w}(t_r)| = c(t-t_r).
\end{equation}

Derived from the Li\'enard-Wiechert potentials
\cite{Griffiths},
the $\bm{E}$ and $\bm{B}$ fields
at the position $\bm{r}$
and time $t$ excited by the point charge
$q$ at the retarded time $t_r$ can be expressed as
\begin{equation}\label{eq:LW_E}
    \bm{E}(\bm{r},t) = \dfrac{q \eta
    [
    (c^2-v^2) \bm{v}' + \bm{\eta} \times (\bm{v}' \times \bm{a})
    ]
    }{4 \pi \varepsilon_0 (\bm{\eta} \cdot \bm{v}')^3}
    ,
\end{equation}
\begin{equation}\label{eq:LW_B}
    \bm{B}(\bm{r},t) = \dfrac{1}{c}
    \bm{\hat{\eta}} \times \bm{E}(\bm{r},t),
\end{equation}
where
$\varepsilon_0$ is the vacuum permittivity,
\begin{equation}
    \bm{\eta} \equiv \bm{r} - \bm{w}(t_r)
\end{equation}
is the position vector pointing from
the retarded position $\bm{w}(t_r)$ to the evaluated position $\bm{r}$,
and its unit vector is $\bm{\hat{\eta}} = \bm{\eta} / \eta$.
The introduced vector $\bm{v}'$ is defined as
\begin{equation}
    \bm{v}' \equiv c \bm{\hat{\eta}} - \bm{v},
\end{equation}
$\bm{v}$ and $\bm{a}$ are the velocity and the acceleration
evaluated at the retarded time $t_r$ of the charge $q$.
The relativistic equations of motion
is needed to close the system,
\begin{equation}\label{eq:drdt2}
    \dfrac{d \bm{r}}{d t} = \bm{v} = \dfrac{\bm{u}}{\gamma},
\end{equation}
\begin{equation}\label{eq:dvdt2}
    \dfrac{d \bm{u}}{d t} = \dfrac{q}{m}
    (\bm{E} + \bm{v} \times \bm{B}),
\end{equation}
where $\gamma = (1-v^2/c^2)^{-1/2} = (1+u^2/c^2)^{1/2}$
is the Lorentz factor,
$m$ is the particle mass.
Note that
the effect of radiation reaction is ignored\cite{Griffiths}.

If we assume a charge $q$ moves with a constant velocity $\bm{v}$,
and it passes the origin $\bm{r}=0$ at time $t=0$,
then $\bm{w} = \bm{v} t$.
The resulting $\bm{E}$ and $\bm{B}$ fields can be written as\cite{Griffiths}:
\begin{equation}\label{eq:LW_E_v0}
    \bm{E}(\bm{r},t) = \dfrac{q}{4 \pi \varepsilon_0}
    \dfrac{1-v^2/c^2}{(1-v^2 \sin^2{\theta} / c^2)^{3/2}}
    \dfrac{\bm{\hat{R}}}{R^2},
\end{equation}
\begin{equation}\label{eq:LW_B_v0}
    \bm{B}(\bm{r},t) = \dfrac{1}{c^2}
    (\bm{v} \times \bm{E}),
\end{equation}
where
$\bm{R} = \bm{r} - \bm{w}(t) = \bm{r} - \bm{v}t$
is the vector from the present location of the charge to $\bm{r}$,
and $\theta$ is the angle between $\bm{R}$ and $\bm{v}$.
Note that the above two equations are reduced version
of the full fields Eq.(\ref{eq:LW_E}) and Eq.(\ref{eq:LW_B}),
because the trajectory of the charge is a straight line,
we do not need to implicitly determine the retarded time anymore.

\subsection{Determination of the Retarded Time}\label{sec:tr}

When the full fields Eq.(\ref{eq:LW_E}) and Eq.(\ref{eq:LW_B})
are considered,
we need to solve the retarded time implicitly.
Therefore, the trajectories $\bm{w}(t)$ of simulated particles
at different time steps are stored.
We use $t_1$ and $t_2$ to denote the range of time,
within which the retarded time lies,
$t_r \in [t_1,t_2]$.
We start to look for the range from a given
maximum range that we are sure $t_r$ must in,
say $t_r \in [t_0,t]$,
where $t_0$ denotes the oldest time that we need to consider,
and $t$ is the present time.
First, if the evaluated position $\bm{r}$ is too close
to the present particle position,
such that $|\bm{r}-\bm{w}(t)|<c \Delta t$,
where $\Delta t$ denotes the simulation timestep,
we can determine
$t_1=t-\Delta t$ and $t_2=t$.
Otherwise, we begin to look for $t_r$ using a bisection method.
The iteration starts at the middle time $t_m$ between
$t_0$ and $t$,
i.e., $t_m = (t_0+t)/2 = t_0 + \delta t = t - \delta t$,
where $\delta t = (t-t_0)/2$ denotes a time interval.
If $c \delta t \leq |\bm{r} - \bm{w}(t_m)|$,
we must have $t_r\in[t_0,t_m]$,
and $(t_0+t_m)/2$ can be assigned as the new $t_m$.
On the contrary, if $c \delta t \geq |\bm{r} - \bm{w}(t_m)|$
we must have $t_r\in[t_m,t]$,
and $(t+t_m)/2$ can be assigned as the new $t_m$.
Then, the new $\delta t$ becomes $(t-t_0)/4$.
The iterations continue until $t_2-t_1=\Delta t$.
Thus, we can approximate these
retarded quantities as $t_r \approx (t_1+t_2)/2$,
$\bm{w}(t_r) \approx [\bm{w}(t_1)+\bm{w}(t_2)]/2$,
$\bm{v}(t_r) \approx [\bm{v}(t_1)+\bm{v}(t_2) ] / 2$,
and $\bm{a}(t_r) \approx [\bm{v}(t_2)-\bm{v}(t_1) ] / \Delta t$.
Note that the history of the particle velocity $\bm{v}(t)$,
or the trajectory in the velocity space,
is also stored,
but the acceleration history is not.

\subsection{Relativistic Particle Pusher}

We apply the relativistic particle pusher
of A. V. Higuera and J. R. Cary \cite{HCpusher},
and follow the procedure described in 
B. Ripperda et al. \cite{Ripperda_2018} to describe the solver
as follows.

1. First half electric field acceleration:
\begin{equation}
    \bm{u}^- = \bm{u}^n + \dfrac{q \Delta t}{2 m} \bm{E}(\bm{r}^{n+1/2}),
\end{equation}
where $n$ denotes the time step.

2. Rotation step:
\begin{equation}
    \bm{u}^+ = s[\bm{u}^- + (\bm{u}^- \cdot \bm{t}) \bm{t} + \bm{u}^- \times \bm{t}].
\end{equation}

3. Second half electric field acceleration:
\begin{equation}
    \bm{u}^{n+1} = \bm{u}^+ + \dfrac{q \Delta t}{2 m} \bm{E}(\bm{r}^{n+1/2})
    + \bm{u}^+ \times \bm{t},
\end{equation}
where note a typo in
\cite{Ripperda_2018}
that the last term should be 
$\bm{u}^+ \times \bm{t}$
instead of 
$\bm{u}^- \times \bm{t}$.

The auxiliary quantities are
$\gamma^- = [1+(\bm{u}^-)^2/c^2]^{1/2}$,
$\bm{\tau} = \bm{B}(\bm{r}^{n+1/2}) q \Delta t / (2 m)$,
$u^* = \bm{u}^- \cdot \bm{\tau} / c$,
$\sigma = (\gamma^-)^2 - \tau^2$,
$\bm{t} = \bm{\tau}/\gamma^+$,
and $s = 1/(1+t^2)$, with
\begin{equation}
    \gamma^+ = \sqrt{\dfrac{\sigma + \sqrt{\sigma^2 + 4(\tau^2+(u^*)^2)}}{2}}.
\end{equation}

\subsection{Frankel's Relativistic Scattering Angle}
Frankel's relativistic scattering angle is defined as follows \cite{PhysRevA.20.2120}:
\begin{equation}\label{eq:Frankel}
\tan{\dfrac{\chi'}{2}} =
\dfrac{q_1 q_2 (c^2 + v_1' v_2')}
{4 \pi \varepsilon_0 E_1' v_1' (v_1'+v_2') b},
\end{equation}
where $b$ is the impact parameter,
the primed quantities are in the center-of-mass frame.
Later we will choose a scenario that
two particles (labeled by 1 and 2)
with the same charge and mass
are going to do a head-on collision
with the same speed, such that the
center-of-mass frame is identical to the lab frame.
Therefore, the primed quantities can be replaced by
quantities in the lab frame,
and $E_1' = E_1 = m_1 c^2 \gamma_1$ is the relativistic energy of particle 1.
Note that the velocities in Eq.(\ref{eq:Frankel})
are the absolute values without signs.

For non-relativistic Coulomb collisions,
Eq.(\ref{eq:Frankel}) reduces to
\begin{equation}
    \tan{\dfrac{\chi}{2}} = \dfrac{b_0}{b},
\end{equation}
where
\begin{equation}
    b_0 = \dfrac{1}{4 \pi \varepsilon_0} \dfrac{q_1 q_2}{\mu (2 v_0)^2},
\end{equation}
where the reduced mass
$\mu = m_1 m_2 / (m_1 + m_2)$.
If we let $b = b_0$,
an exact 90 degree scattering is expected for non-relativistic $v_0$.

\section{Simulation}
\subsection{Test under Constant Velocity}\label{sec:3.1}

In this section,
a test under the constant velocity assumption
is carried out,
comparing using the full fields Eq.(\ref{eq:LW_E}) and Eq.(\ref{eq:LW_B})
with the implicit $t_r$ determination
as described in Sec.\ref{sec:tr}
to the reduced form of Eq.(\ref{eq:LW_E_v0}) and Eq.(\ref{eq:LW_B_v0}),
such that the computation of $t_r$ and $\bm{\eta}$
can be validated, and the accuracy can be evaluated.
The basic parameters used in the test are described below.
$N_{tmax}=1000$ is the max number of time steps
stored for the particle trajectory,
$\Delta t=1$ ns is the timestep,
$v_0=0.01c$ m/s is the constant velocity along $z$,
and the charge is assumed to be a unit charge $e$. Next, we begin to compute the fields at different spatial positions
on the $y=0$ plane
from $x=1$ to 200 m
and $z=1$ to 100 m
at the present time
$t= N_{tmax} \Delta t=1$ $\mu$s.
The number of bins used to compute the field data is
$200 \times 100$.
Note that the charge moves from
the origin to $\bm{w}=(0,0,v_0 t)$, i.e., $z \approx 3$ m,
and the speed of light travels about 300 m during 1 $\mu$s,
meaning the farthest fields established by the charge in space
have reached about 300 m.

\begin{figure}[!ht]
\centering
(a)\includegraphics[width=0.4\textwidth]{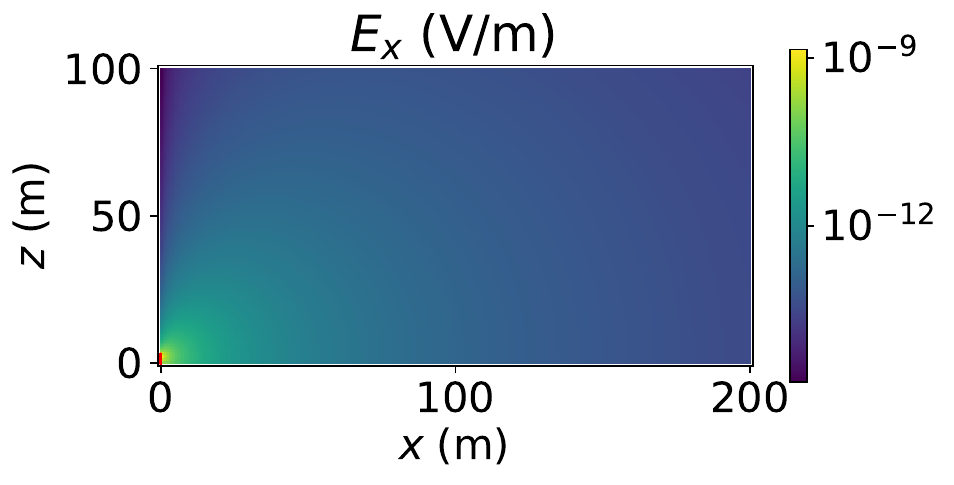}\\
(b)\includegraphics[width=0.4\textwidth]{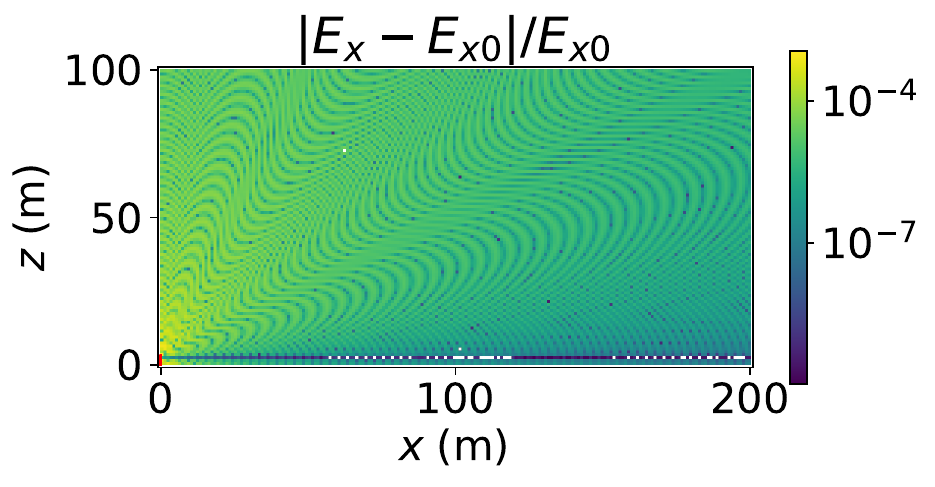}
\caption{
(a) The simulated $E_x$ field.
(b) The relative error of $E_x$.
}
\label{fig:Ex_Exerr}
\end{figure}

The simulation results of $E_x$ are plotted in
Fig.\ref{fig:Ex_Exerr} (a),
which is the field computed from
Eq.(\ref{eq:LW_E}) and Eq.(\ref{eq:LW_B})
solving the retarded time implicitly.
The simulation result obtained by solving
Eq.(\ref{eq:LW_E_v0}) and Eq.(\ref{eq:LW_B_v0})
is denoted by $E_{x0}$,
which is very close to $E_x$,
thus not shown,
but the relative error is plotted in
Fig.\ref{fig:Ex_Exerr} (b).
We can see the computation leads to maximum
relative error below $10^{-3}$.
Note that the trajectory of the charge is also marked
in the figure by the short red line
ranging from $z=0$ to $3$ m.
Those white dots or lines in Fig.\ref{fig:Ex_Exerr} (b)
indicate relative errors that are close to zero,
such that no values can be obtained in the log scale.

\begin{figure}[h!]
\centering
(a)\includegraphics[width=0.4\textwidth]{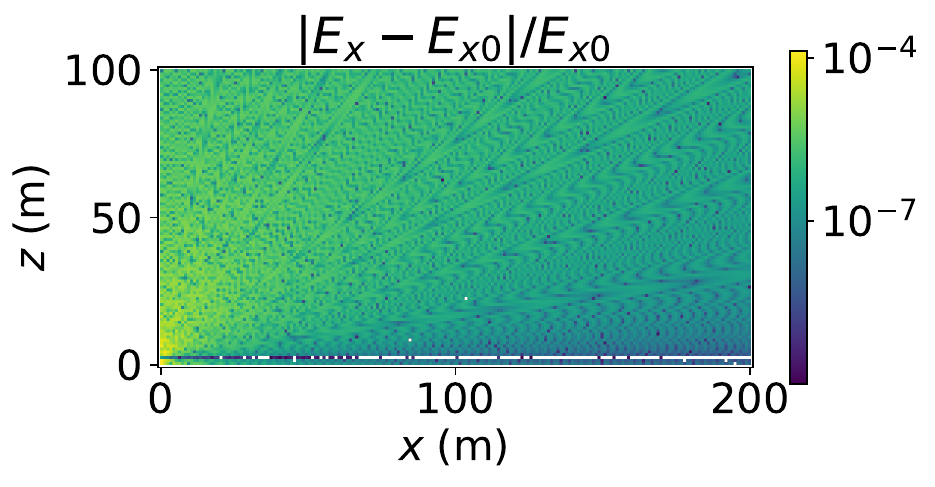}\\
(b)\includegraphics[width=0.4\textwidth]{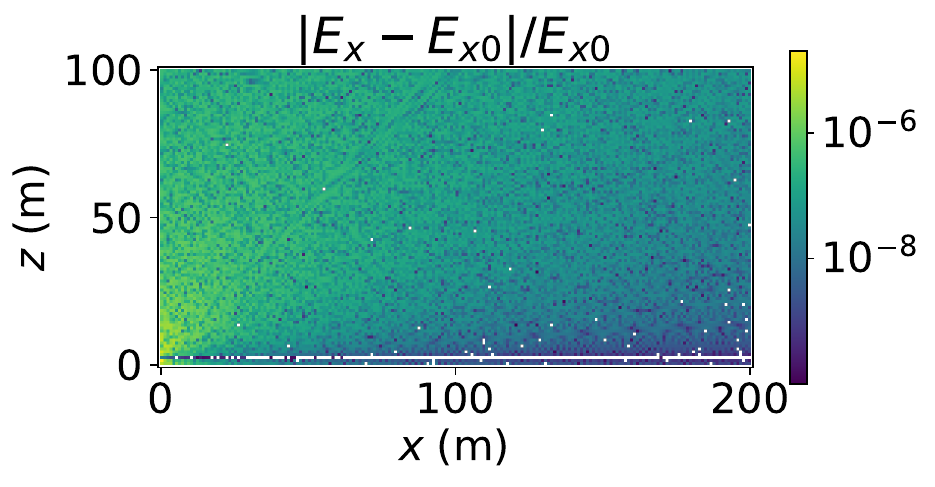}
\caption{
The relative error of $E_x$ with
different timesteps, $\Delta t=0.1$ ns (a)
and 0.01 ns (b).
}
\label{fig:Exerr_dt}
\end{figure}

\begin{figure}[h!]
\centering
(a)\includegraphics[width=0.4\textwidth]{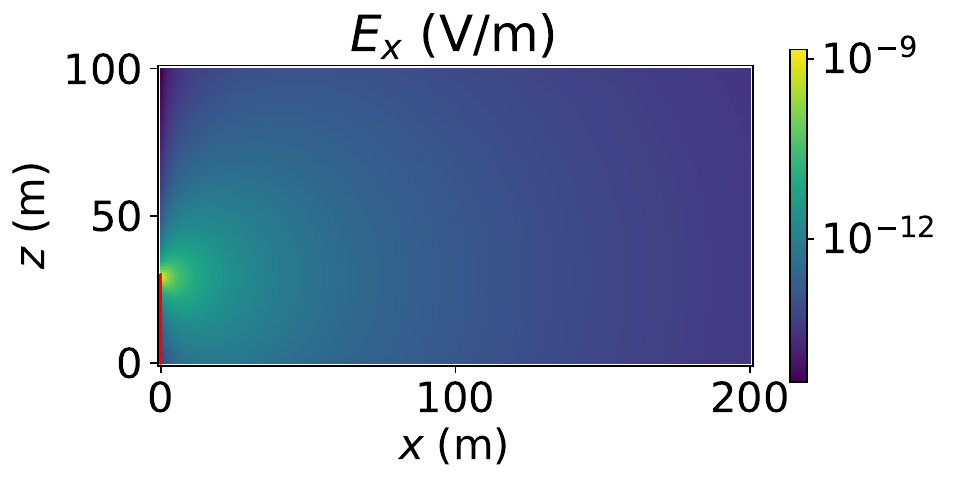}\\
(b)\includegraphics[width=0.4\textwidth]{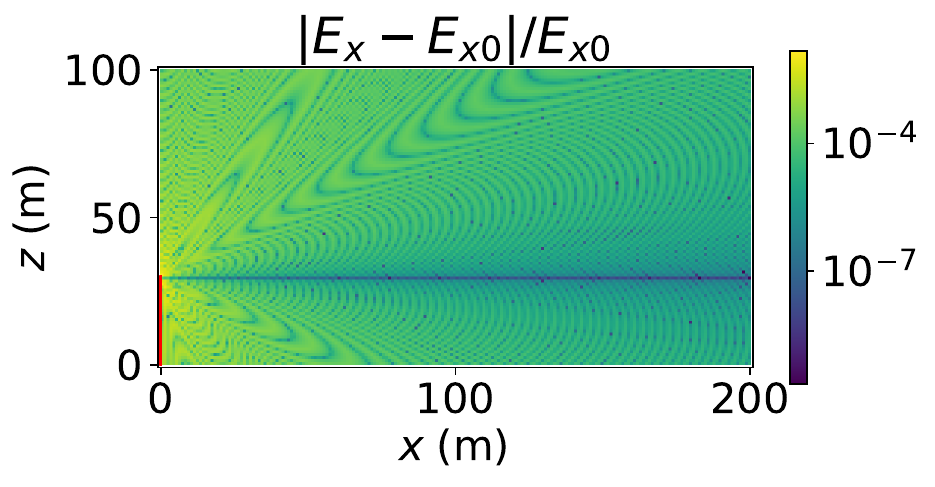}
\caption{
The simulation results of $E_x$ with
$v_0=0.1c$.
}
\label{fig:Ex_Exerr_v0_0d1}
\end{figure}

Then, the timestep $\Delta t$ is tested,
by decreasing it to be 0.1 ns and 0.01 ns,
thus $N_{tmax}$ is increased to be 10000 and 100000,
accordingly.
The relative errors of $E_x$ are presented in Fig.\ref{fig:Exerr_dt}.
As we can see, decreasing $\Delta t$ leads
a higher resolution of the charge's stored trajectory,
thus higher simulation accuracy.

Next, changing back to $\Delta t = 1$ ns,
the charge velocity is increased to $v_0=0.1c$ from $0.01c$.
The simulation results are shown in Fig.\ref{fig:Ex_Exerr_v0_0d1}.
We can see that the charge's trajectory becomes longer,
and the relative error increases a little bit.

\begin{figure}[!ht]
\centering
\includegraphics[width=0.24\textwidth]{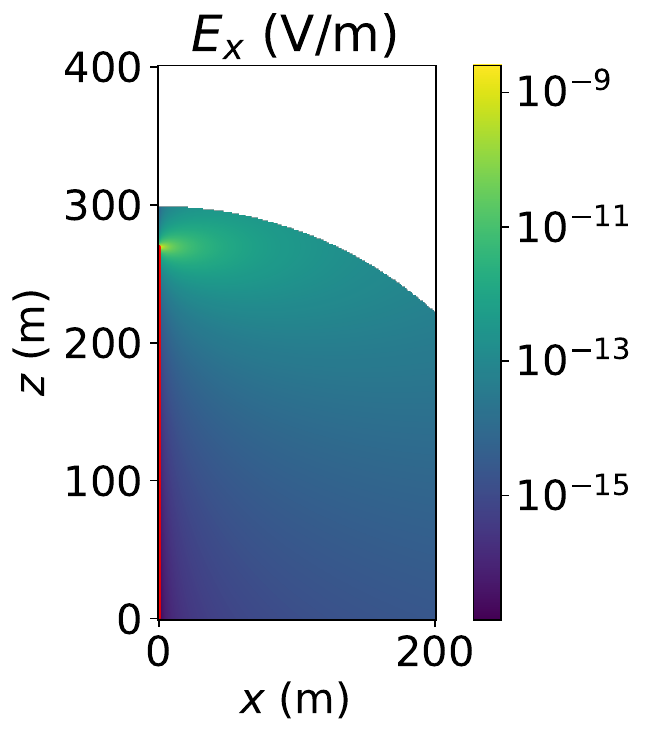}
\includegraphics[width=0.24\textwidth]{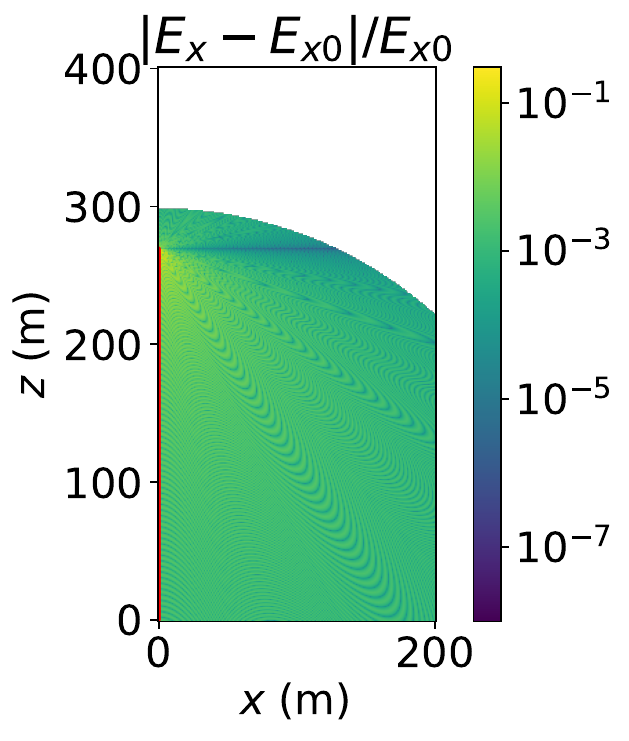}
\includegraphics[width=0.24\textwidth]{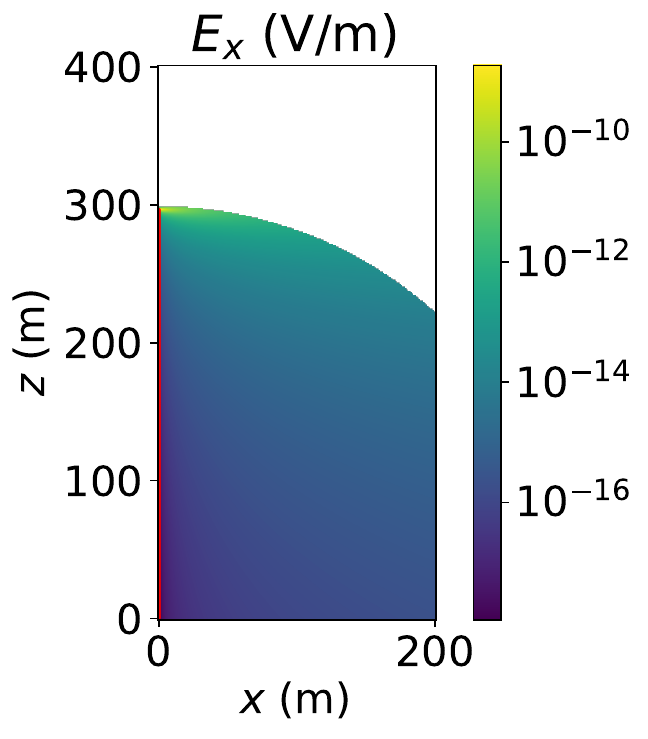}
\includegraphics[width=0.24\textwidth]{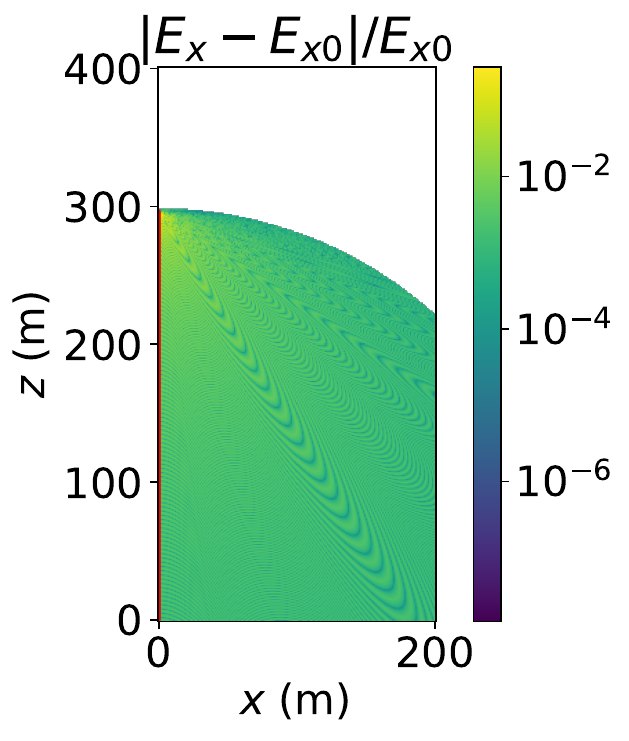}
\caption{
The simulation results of $E_x$ with
$v_0=0.9c$ (top two plots) and $0.99c$ (bottom two plots).
}
\label{fig:Ex_Exerr_v0_0d9}
\end{figure}

Further more, we increase $v_0$ to be $0.9c$ ($\gamma \approx 2.3$)
and $0.99c$ ($\gamma \approx 7.1$),
which becomes relativistic.
The simulation results are shown in Fig.\ref{fig:Ex_Exerr_v0_0d9}.
Note that this time the range of $z$ is enlarged to be 400
covering the whole particle trajectory.
We can see the round shape indicating the region where
the fields have been established,
while the region of blank has no fields arrived yet.
The relative error becomes even larger,
and higher resolution of the trajectory is needed.

\subsection{Relativistic Binary Collision}\label{sec:3.2}
\subsubsection{The Simulation Setup}

\begin{figure}[!ht]
\centering
\includegraphics[width=0.5\textwidth]{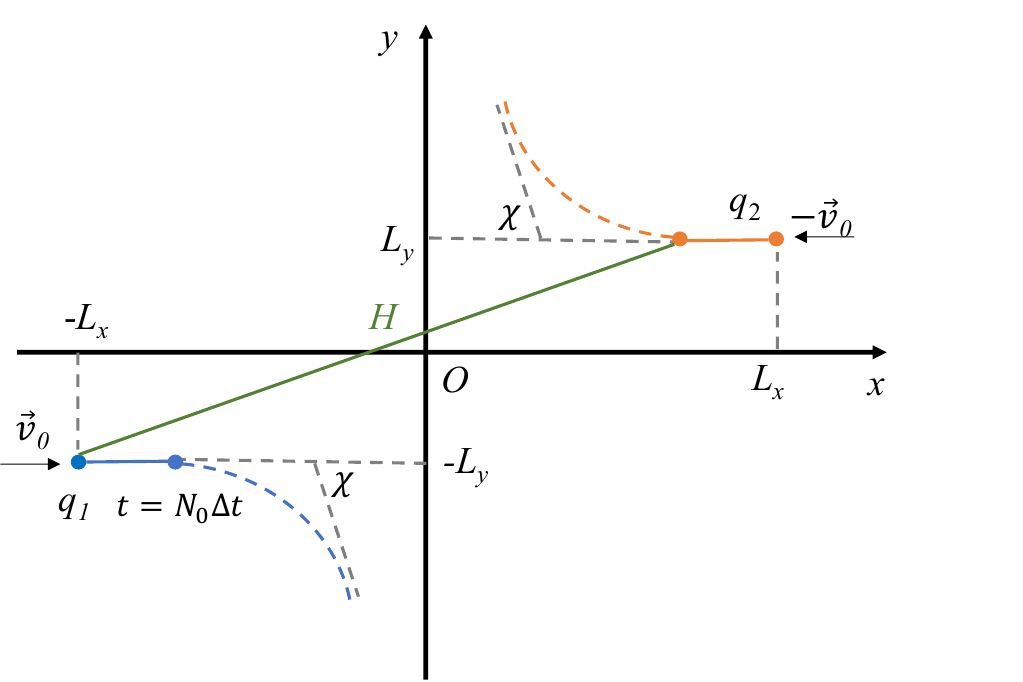}
\caption{
Simulation setup of the binary collision.
}
\label{fig:diagram}
\end{figure}

In this section, we start to use the EM-PP model
to study the relativistic binary collision.
The simulation setup is illustrated in Fig.\ref{fig:diagram}.
We have two charges in vacuum,
$q_1$ is located at $(-L_x,-L_y)$
and $q_2$ is located at $(L_x,L_y)$
at time $t=0$.
The commonly used impact parameter is thus $b=2L_y$.
Let us assume the two charges have the same initial velocity $v_0$
in the $\pm x$ direction.
Then, the two charges will collide with each other over time,
and result in a scattering angle $\chi$.

Because the fields that generated by one charge
need to spend some time to travel
to the other charge at the speed of light $c$,
we need to provide particle trajectories
that store enough ``old'' positions for
computing the fields at the retarded times.
Thus, we let the two charges move with their initial
velocities $N_0$ time steps,
assuming that $L_x$ is chosen large enough
such that the two charges are not interacting each other much
until $t=N_0 \Delta t$.
Obviously, we do not want $N_0$ and $L_x$ be too large
for saving computations,
the minimal $N_0$ needed for a certain $L_x$ can be derived as follows.

The distance $H$ labeled in Fig.\ref{fig:diagram}
connects the position of $q_1$ at $t=0$ and that of $q_2$ at $t=N_0 \Delta t$,
we want to make sure that $H \leq c N_0 \Delta t$, i.e.,
\begin{equation}
    \sqrt{(2 L_x - N_0 \Delta t v_0)^2 + (2 L_y)^2} \leq c N_0 \Delta t.
\end{equation}
Applying the quadratic formula leads to
\begin{equation}\label{eq:N0}
    N_0 \geq \dfrac{-B + \sqrt{B^2 - 4 A C}}{2 A},
\end{equation}
where
$A = (c^2 - v_0^2)\Delta t^2$,
$B = 4 L_x \Delta t v_0$,
$C = - 4 (L_x^2 + L_y^2)$,
and we have dropped the solution with the negative sign.
We can therefore choose the minimal integer of $N_0$ that satisfies Eq.(\ref{eq:N0}).
After $N_0$ is determined,
we set $N_{tmax}=20N_0$,
which is found to be large enough.
The simulation will start with a certain
$\Delta t$, $v_0$, $L_x$,
and $b=b_0$ is applied first,
such that a 90 degree scattering is expected for
$v_0 \ll c$.
In addition, for simplicity,
we would like to choose $q_1=q_2=e$,
and the particle mass $m_1=m_2=m_e$,
thus the two particles are positrons,
and the center-of-mass frame coincides with the lab frame.
Note that from $t=0$ to $N_0 \Delta t$,
we assume the $y$ of two particles are fixed at
$\pm L_y$,
and their velocities are $\pm v_0$,
so their $x$ positions are computed accordingly.
Then the fields generated by two particles are calculated, and acceleration and motion are updated correspondingly.

\subsubsection{Simulation Results}
When $v_0$ is much smaller than $c$,
if we let $b=b_0$,
the scattering angle should be $\pi/2$.
We first carry out tests with $v_0/c=0.1$
and $b=b_0\approx 1.409\times10^{-13}$ m.
By varying $\Delta t$ and $L_x$,
we can obtain converging $\chi$,
as shown in Tab.\ref{tab:test_0d1c}.
Smaller $\Delta t$ and larger $L_x$ lead to
more accurate results,
the scattering angle is about
$\chi_1/\pi \approx 0.498064$,
which is very close to 0.5, the non-relativistic value.
The subscript ``1'' of $\chi_1$ corresponds to the
solution using the full fields
Eq.(\ref{eq:LW_E}) and Eq.(\ref{eq:LW_B}).
Similarly, tests with $v_0/c=$0.25, 0.45, 0.6, 0.75, 0.9, 0.99, 0.999, and 0.9999
are simulated and shown from Tab.\ref{tab:test_0d25c} to Tab.\ref{tab:test_0d9999c}
in the appendix.
We present these tables to show the convergence of each test.

\begin{table}[!ht]
\centering
\begin{tabular}{ccrc} 
  \hline
  $\log_{10}\Delta t$ (s) & $L_x/L_y$ & $N_0$ & $\chi_1/\pi$ \\ 
  \hline
  -22 & 1000  & 4273    & 0.497730 \\
  -23 & 1000  & 42726   & 0.497697 \\
  -24 & 1000  & 427257  & 0.497697 \\
  -24 & 2000  & 854513  & 0.497884 \\
  -24 & 4000  & 1709025 & 0.497978 \\
  -23 & 4000  & 170903  & 0.497982 \\
  -23 & 8000  & 341805  & 0.498029 \\
  -24 & 8000  & 3418050 & 0.498025 \\
  -23 & 16000 & 683610  & 0.498052 \\
  -23 & 32000 & 1367220 & 0.498064 \\
  \hline
\end{tabular}
\caption{Tests with $v_0/c=0.1$ and $b=b_0\approx 1.409\times10^{-13}$ m.}
\label{tab:test_0d1c}
\end{table}

\begin{figure}[!ht]
\centering
(a)\includegraphics[width=0.33\textwidth]{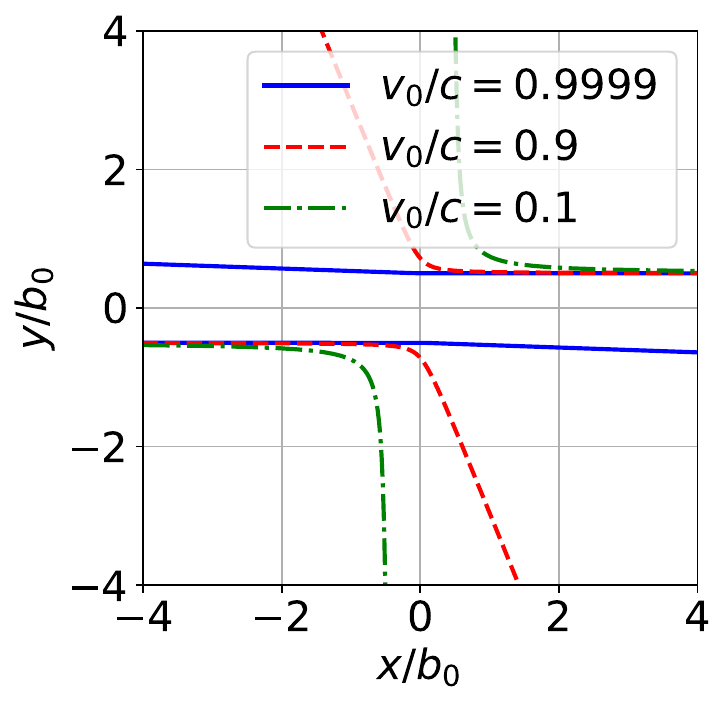}\\
(b)\includegraphics[width=0.4\textwidth]{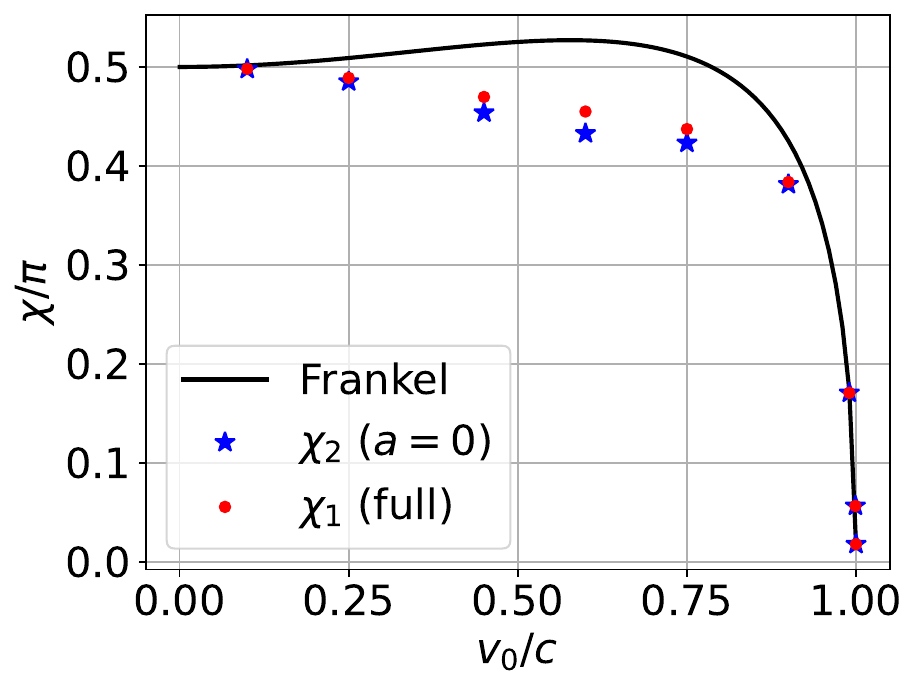}\\
(c)\includegraphics[width=0.4\textwidth]{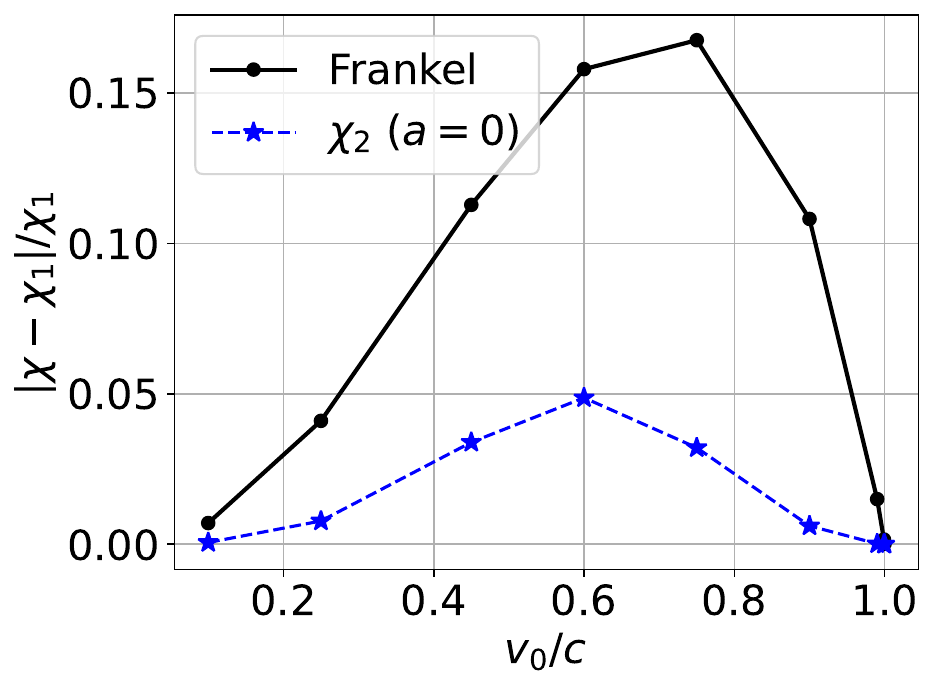}
\caption{
(a) The simulated binary collision trajectories at different velocities.
(b) Scattering angle over velocity of different solutions.
(c) Relative errors of Frankel's angle and $\chi_2$
compared to $\chi_1$.
}
\label{fig:w1_angle}
\end{figure}

The binary collision process can be seen from Fig.\ref{fig:w1_angle} (a).
The scattering angle as a function of
velocity $v_0/c$ is plotted in 
Fig.\ref{fig:w1_angle} (b),
where $\chi_2$ denotes the simulated results under
the assumption $\bm{a}=0$.
In addition, Frenkel's scattering angle formula
Eq.(\ref{eq:Frankel}) is plotted too.
As we can see from Fig.\ref{fig:w1_angle} (b),
as well as the relative error plot in Fig.\ref{fig:w1_angle} (c),
there is a region between $v_0/c=0.1$ to 0.99
that Frankel's formula
leads to less accurate results.
The maximum relative error is about 17\%
at $v_0/c=0.75$.
The zero-acceleration simplification ($\chi_2$)
leads to less accurate results
between $v_0/c = 0.1$ to 0.9,
and it is overall more accurate than Frankel's formula.
The maximum relative error is about 5\%
at $v_0/c=0.6$.

Recalling the derivation process of Frankel's
relativistic scattering angle in \cite{PhysRevA.20.2120},
the relatively large errors are due to
(1) Frankel actually applied Eq.(\ref{eq:LW_E_v0})
and Eq.(\ref{eq:LW_B_v0}),
thus not only the effect of non-zero $\bm{a}$
is ignored,
but also particles are assumed to have constant velocities;
(2) Frankel used the impulse approximation,
such that the collision process is
assumed to be so brief that the particles
do not significantly change their positions
during the interaction.
Thus, it is only valid for small-angle scatterings.
Comparing the results of $\chi_2$ to $\chi_1$,
we can see that the acceleration term
in Eq.(\ref{eq:LW_E})
is important only when the velocity is relativistic
and there is a large scattering,
such that $\bm{a}$ is sufficiently greater than zero
to play a role.

\section{Conclusion}\label{sec:conclusion}

In this paper,
an electromagnetic particle-particle
(EM-PP) model is proposed,
by solving the electric and
magnetic fields derived
from the
Li\'enard-Wiechert potentials
of point charge.
It is shown that the
retarded time can be
computed implicitly
using a bisection method,
and the relativistic particle
equations of motion
can be solved efficiently using
Higuera's method,
such that the EM-PP model
is validated
by a constant-velocity test
and relativistic binary collisions.
In addition,
Frankel's relativistic
scattering angle formula
is compared to the
baseline results obtained
by EM-PP.
It is found that
for scattering angle greater
than about 0.2$\pi$,
Frankel's formula begins to
become less accurate,
since it was developed based on
the small-angle scattering assumption,
and the maximum relative error
observed is about 17\%
at speed $0.75 c$
and scattering angle
0.43$\pi$.
By the way,
it is shown that
a simplified version 
of the EM-PP model
under the zero-acceleration assumption
can also achieve relatively
good accuracy.
Under the relativistic binary collision
tests,
only 5\% maximum relative error is
observed
at speed about 0.6$c$
and angle 0.45$\pi$.
In the future,
the EM-PP model
will be applied to
other plasma applications involving
more relativistic particles.

\section*{Acknowledgment}

The authors acknowledge the support from
National Natural Science Foundation of China
(Grant No. 5247120164).


\appendix

\section*{Appendix}

Tests with $v_0/c=$0.25, 0.45, 0.6, 0.75,
0.9, 0.99, 0.999, and 0.9999
are simulated and shown from Tab.\ref{tab:test_0d25c} to Tab.\ref{tab:test_0d9999c},
indicating the convergence of each test.

\begin{table}[!ht]
\centering
\begin{tabular}{ccrc} 
  \hline
  $\log_{10}\Delta t$ (s) & $L_x/L_y$ & $N_0$ & $\chi_1/\pi$ \\ 
  \hline
  -23 & 32000 & 192505  & 0.488885 \\
  -24 & 32000 & 1925046 & 0.488964 \\
  -24 & 64000 & 3850092 & 0.488970 \\
  \hline
\end{tabular}
\caption{Tests with $v_0/c=0.25$ and $b=b_0\approx 2.254\times10^{-14}$ m.}
\label{tab:test_0d25c}
\end{table}

\begin{table}[!ht]
\centering
\begin{tabular}{ccrc} 
  \hline
  $\log_{10}\Delta t$ (s) & $L_x/L_y$ & $N_0$ & $\chi_1/\pi$ \\ 
  \hline
  -23 & 64000 & 102440  & 0.471405 \\
  -24 & 64000 & 1024397 & 0.469750 \\
  -25 & 32000 & 5121982 & 0.469679 \\
  \hline
\end{tabular}
\caption{Tests with $v_0/c=0.45$ and $b=b_0\approx 6.958\times10^{-15}$ m.}
\label{tab:test_0d45c}
\end{table}

\begin{table}[!ht]
\centering
\begin{tabular}{ccrc} 
  \hline
  $\log_{10}\Delta t$ (s) & $L_x/L_y$ & $N_0$ & $\chi_1/\pi$ \\ 
  \hline
  -24 & 64000 & 522203  & 0.455028 \\
  -25 & 32000 & 2611011 & 0.454960 \\
  -25 & 64000 & 5222021 & 0.454961 \\
  \hline
\end{tabular}
\caption{Tests with $v_0/c=0.6$ and $b=b_0\approx 3.914\times10^{-15}$ m.}
\label{tab:test_0d6c}
\end{table}

\begin{table}[!ht]
\centering
\begin{tabular}{ccrc} 
  \hline
  $\log_{10}\Delta t$ (s) & $L_x/L_y$ & $N_0$ & $\chi_1/\pi$ \\ 
  \hline
  -24 & 32000 & 152782  & 0.435988 \\
  -25 & 32000 & 1527815 & 0.437213 \\
  -25 & 64000 & 3055629 & 0.437225 \\
  \hline
\end{tabular}
\caption{Tests with $v_0/c=0.75$ and $b=b_0\approx 2.505\times10^{-15}$ m.}
\label{tab:test_0d75c}
\end{table}

\begin{table}[!ht]
\centering
\begin{tabular}{ccrc} 
  \hline
  $\log_{10}\Delta t$ (s) & $L_x/L_y$ & $N_0$ & $\chi_1/\pi$ \\ 
  \hline
  -25 & 32000 &  977221 & 0.383702 \\
  -25 & 64000 & 1954441 & 0.383562 \\
  -26 & 16000 & 4886102 & 0.383745 \\
  \hline
\end{tabular}
\caption{Tests with $v_0/c=0.9$ and $b=b_0\approx 1.739\times10^{-15}$ m.}
\label{tab:test_0d9c}
\end{table}

\begin{table}[!ht]
\centering
\begin{tabular}{ccrc} 
  \hline
  $\log_{10}\Delta t$ (s) & $L_x/L_y$ & $N_0$ & $\chi_1/\pi$ \\ 
  \hline
  -25 & 32000 & 771095  & 0.170914 \\
  -26 & 16000 & 3855473 & 0.170846 \\
  -26 & 32000 & 7710946 & 0.170843 \\
  \hline
\end{tabular}
\caption{Tests with $v_0/c=0.99$ and $b=b_0\approx 1.436\times10^{-15}$ m.}
\label{tab:test_0d99c}
\end{table}

\begin{table}[!ht]
\centering
\begin{tabular}{ccrc} 
  \hline
  $\log_{10}\Delta t$ (s) & $L_x/L_y$ & $N_0$ & $\chi_1/\pi$ \\ 
  \hline
  -25 & 32000 & 753855  & 0.0562335 \\
  -26 & 16000 & 3769271 & 0.0566305 \\
  -26 & 32000 & 7538542 & 0.0566306 \\
  \hline
\end{tabular}
\caption{Tests with $v_0/c=0.999$ and $b=b_0\approx 1.412\times10^{-15}$ m.}
\label{tab:test_0d999c}
\end{table}

\begin{table}[!ht]
\centering
\begin{tabular}{ccrc} 
  \hline
  $\log_{10}\Delta t$ (s) & $L_x/L_y$ & $N_0$ & $\chi_1/\pi$ \\ 
  \hline
  -25 & 32000 & 752159  & 0.0111283 \\
  -26 & 16000 & 3760795 & 0.0179951 \\
  -26 & 32000 & 7521590 & 0.0179960 \\
  -27 & 8000  & 4700995 & 0.0179076 \\
  \hline
\end{tabular}
\caption{Tests with $v_0/c=0.9999$ and $b=b_0\approx 1.409\times10^{-15}$ m.}
\label{tab:test_0d9999c}
\end{table}

\section*{Data Availability}
The data that support the findings of this study are available from the corresponding author upon reasonable request.

\section*{References}
\printbibliography[heading=none]

\end{document}